\documentclass[intlimits,twoside,a4paper]{article}

\usepackage{graphicx}        

\usepackage[T2A,T1]{fontenc} 
\usepackage[cp1251]{inputenc}

\usepackage{cmpj3}

\issue{2021}{24}{3}{33602}
\doinumber{10.5488/CMP.24.33602}

\title[A simple second order thermodynamic perturbation theory for associating fluids]{A simple second order thermodynamic perturbation theory for associating fluids
}
\author[B. D. Marshall]{B. D. Marshall\orcid{0000-0002-3079-5946}
\thanks{\email{Bennett.d.marshall@exxonmobil.com}}}
\address{ExxonMobil Research and Engineering, 1545 Rt. 22 East, Annandale, NJ 08801}
\Keywords{thermodynamics, classical statistical mechanics, self-assembly}
\date{Received June  13, 2021, in final form August 12, 2021}

\begin{document}
\maketitle

\begin{abstract}

An approximation within Wertheim's second order perturbation theory is proposed which allows for the development of a general solution for pure component fluids with an arbitrary number and functionality of association sites. The solution is closed, concise and general for all second order effects such as ring formation, steric hindrance and hydrogen bond cooperativity. The approach is validated by comparison to hydrogen bond structure data for liquid water. 
\printkeywords

\end{abstract}
%
%

\section{Introduction}

 This paper is in honor of Dr. Yurij Kalyuzhnyi and his many outstanding theoretical contributions to the thermodynamics and structure of associating fluids. 

 The thermodynamics and liquid structure of associating fluids pervades many areas of physical, chemical, materials and engineering research and application. Associating fluids interact with highly directional potentials with a limited valence. In materials research, patchy colloids \cite{1} represent possible avenues for self-assembled materials as well as primitive models for proteins \cite{2}. The thermodynamics of molecular fluids such as water, alcohols, carboxylic acids, etc.,  are dominated by the effects of hydrogen bonds, an association interaction. 

 Wertheim's multi-density statistical mechanics \cite{3,4} has become a cornerstone in the theoretical description of associating fluids. Dr. Kalyuzhnyi has extensively worked within this multi-density formalism to develop a series of integral equation theories \cite{5,6,7} in order to predict the liquid structure of associating fluids. Thermodynamic perturbation theories (TPT) within this multi-density formalism have been widely adopted within engineering equations of state \cite{8} for hydrogen bonding fluids as well as primitive models \cite{9} for patchy colloids. 

 TPT is typically applied at first order in perturbation, TPT1. TPT1 treats each association bond independently. This results in independent bonding probabilities for each association site allowing for a simple, general solution of TPT1 for fluids of arbitrary complexity. This simplicity comes at a price. Contributions to the free energy which arise through the interaction of three associating units (colloids, molecules) such as steric hindrance, hydrogen bond cooperativity, ring formation, and multiple bonding of a single association site cannot be described in TPT1. To include these effects one must go to second order in perturbation, TPT2. Unfortunately, in TPT2 the simplicity vanishes and a general solution to the theory cannot be stated in concise form. TPT2 theories necessarily lead to the evaluation of recursion relations~\cite{10,11} in fluids with > 2 association sites, which would be difficult to generalize. This complexity limits the widespread applicability of TPT2.

 In this paper a ``middle ground'' between TPT1 and TPT2 is proposed. An approach which allows for the inclusion of second order effects, but retains a portion of the simplicity and generality of TPT1.

\section{Simplified TPT2}

In this section a pure fluid of $N$ associating spheres in a volume $V$ at a temperature $T$ is considered. The spheres contain a set $\Gamma =\left\{A,B,C,\ldots \right\}$ of association sites labelled with capital letters. The potential of interaction between sphere 1 and sphere 2 is given by

\begin{equation} \label{1} \phi \left(12\right)=\phi _{hs}^{} \left(r_{12} \right)+\sum _{A\in \Gamma }\sum _{B\in \Gamma }\phi _{AB}^{} \left(12\right) .  \end{equation} 

The distance between the centers of the molecules is ${r_{12}}$ and the notation (\ref{1}) represents the position and orientation of molecule 1. The term \textit{$\varphi_{hs}$} is the pair potential of the spherically symmetric hard sphere reference fluid and $\phi _{AB}$ is the potential of interaction between sites \textit{A} and \textit{B}. 

 In Wertheim's \cite{3,4} formalism, the density of spheres \textit{$\rho_\alpha$} bonded at the set of sites $\alpha $ plays a central role. Based on these set densities, Wertheim further defined the density parameters $\sigma_\alpha$ which act as field points in the multi-density cluster expansion

\begin{equation} \label{2} \sigma _{\alpha }^{} =\sum _{\gamma \subset \alpha }\rho _{\alpha }^{} .  \end{equation} 

The monomer density is $\rho_{o}$= $\sigma_{o}$, and the total density is given by $\rho $ = $\sigma_{\Gamma}$. The densities are related to the fundamental graph sum ${c_{o}}$ through the following relation

\begin{equation} \label{3} \frac{\rho _{\gamma }^{} }{\rho _{o}^{} } =\sum _{P\left(\gamma \right)=\left\{\tau \right\}}\prod _{\tau }c_{\tau }^{} \,,  \end{equation} 
where
\begin{equation} \label{4} c_{\gamma }^{} =\frac{\partial }{\partial \sigma _{\Gamma -\gamma }^{} } \frac{\Delta c^{\left(o\right)} }{V} \; ;\quad \gamma \ne \emptyset. \end{equation}

In second order perturbation theory (TPT2) the graph sum is decomposed as

\begin{equation} \label{5} c^{\left(o\right)} -c_{\text{ref}}^{(o)} =\Delta c_{}^{\left(o\right)} =\Delta c_{I}^{\left(o\right)} +\Delta c_{II}^{\left(o\right)}.  \end{equation} 
$c_{\text{ref}}^{(o)}$ is the contribution for the non-associating reference fluid, $\Delta c_{I}^{\left(o\right)} $is the first order contribution which contains graphs with a single association bond, and $\Delta c_{II}^{\left(o\right)} $ contains second order graphs with two association bonds. From equations (\ref{4})--(\ref{5}) 

\begin{equation} \label{6} \begin{array}{l} {c_{\alpha } =\frac{\partial \Delta c_{I}^{\left(o\right)} }{\partial \sigma _{\Gamma -\alpha } } =0\quad{\text{for}}\quad n\left(\alpha \right)>1}, \\ {} \\ {c_{\alpha } =\frac{\partial \Delta c_{II}^{\left(o\right)} }{\partial \sigma _{\Gamma -\alpha } } =0\quad {\text{for}}\quad n\left(\alpha \right)>2}. \end{array} \end{equation} 

 A major obstacle in the development of a general TPT2 solution is the general enumeration of the site densities. Consider spheres bonded with a set of 4 association sites $\alpha =\left\{A,B,C,D\right\}$

%
\begin{eqnarray} \label{7)}  &&\frac{\rho _{ABCD} }{\rho _{o} } =c_{A} c_{B} c_{C} c_{D} +c_{AB} c_{C} c_{D} +c_{BC} c_{A} c_{D} +c_{CD} c_{A} c_{B} +c_{AC} c_{B} c_{D} +c_{BD} c_{A} c_{C}  \nonumber\\ 
&&{+c_{AD} c_{B} c_{C} +c_{AB} c_{CD} +c_{AC} c_{BD} +c_{AD} c_{BC} }.  \end{eqnarray}

As the number of sites increases in the set \textit{$\alpha $}, this enumeration becomes increasingly complex. In addition, there is no general analytical way to evaluate these sums. 

 To develop a simpler analytical TPT2 theory, we assume that second order corrections governed by the $c_{AB}$ will be small compared to the first order corrections given by $c_{A}$. Hence, contributions to {$\rho_\alpha $} which contain products for two or more second order terms $c_{AB}c_{CD}$ can be neglected. In other words, we make the following assumption,

\begin{equation} \label{8} c_{\alpha } c_{\beta } =0 \quad \text{for}\quad n\left(\alpha \right)=n\left(\beta \right)=2.
\end{equation} 

 Theoretically, the simplification in equation~(\ref{8}) results in the limitation that each associating sphere can only participate in a single second order interaction at a time. There can be multiple pairs of association sites which exhibit second order properties, but any associated state of the sphere is limited to a single second order interaction. Combining equations~(\ref{3}) and (\ref{8})

\begin{equation} \label{9} \frac{\rho _{\alpha } }{\rho _{o} } =\prod _{A\in \alpha }c_{A}  +\sum _{AB\in \alpha }c_{AB} \prod _{D\in \alpha -AB}c_{D}.    \end{equation} 

From equation~(\ref{3}) 

\begin{equation} \label{10} c_{A} =\frac{\rho _{A} }{\rho _{o} };\quad c_{AB} =\frac{\rho _{AB} }{\rho _{o} } -\frac{\rho _{A} }{\rho _{o} } \frac{\rho _{B} }{\rho _{o} }.  \end{equation} 

Combining (\ref{9}) and (\ref{10})

%
\begin{eqnarray} \label{11}  
\frac{\rho _{\alpha } }{\rho _{o} } &=&\prod _{A\in \alpha }\frac{\rho _{A} }{\rho _{o} }  +\sum _{AB\in \alpha }\left(\frac{\rho _{AB} }{\rho _{o} } -\frac{\rho _{A} }{\rho _{o} } \frac{\rho _{B} }{\rho _{o} } \right)\prod _{D\in \alpha -AB}\frac{\rho _{D} }{\rho _{o} }    \nonumber\\ 
&=&\prod _{A\in \alpha }\frac{\rho _{A} }{\rho _{o} }  +\sum _{AB\in \alpha }\frac{\rho _{AB} }{\rho _{o} } \prod _{D\in \alpha -AB}\frac{\rho _{D} }{\rho _{o} }  -\sum _{AB\in \alpha }\frac{\rho _{A} }{\rho _{o} } \frac{\rho _{B} }{\rho _{o} } \prod _{D\in \alpha -AB}\frac{\rho _{D} }{\rho _{o} }     \nonumber\\  
&=&\sum _{AB\in \alpha }\frac{\rho _{AB} }{\rho _{o} } \prod _{D\in \alpha -AB}\frac{\rho _{D} }{\rho _{o} }  -\left(\lambda _{\alpha } -1\right) \prod _{A\in \alpha }\frac{\rho _{A} }{\rho _{o} }  .  \end{eqnarray}

The constant \textit{$\lambda_\alpha $} represents the numbers of pairs of sites in \textit{$\alpha $}. Equation (\ref{11}) can be rewritten with the aid of site operators~\cite{4}

\begin{equation} \label{12} \hat{\sigma }_{\alpha } =\sum _{AB\in \alpha }\hat{\sigma }_{AB} \prod _{D\in \alpha -AB}\hat{\sigma }_{D}  -\left(\lambda _{\alpha } -1\right) \prod _{D\in \alpha }\hat{\sigma }_{D}  =\left(\sum _{AB\in \alpha }\frac{\hat{\sigma }_{AB} }{\hat{\sigma }_{A} \hat{\sigma }_{B} } -\left(\lambda _{\alpha } -1\right) \right)\prod _{D\in \alpha }\hat{\sigma }_{D},   \end{equation} 
where $\hat{\sigma }_{\alpha } =\sigma _{\alpha } /\rho _{o} $ and

\begin{equation} \label{13} \frac{\sigma _{A} }{\rho _{o} } =1+c_{A} ;\quad \frac{\sigma _{AB} }{\rho _{o} } =\left(1+c_{A} \right)\left(1+c_{B} \right)+c_{AB}.  
\end{equation} 

We simplify equation (\ref{12})

\begin{eqnarray} \label{14} {\frac{\sigma _{\alpha } }{\rho _{o} } =\left[\sum _{AB\in \alpha }\left(1+\frac{c_{AB} }{\left(1+c_{A} \right)\left(1+c_{B} \right)} \right)-\left(\lambda _{\alpha } -1\right) \right]\prod _{D\in \alpha }\frac{\sigma _{D} }{\rho _{o} }  } \nonumber\\
 =\left(1+\sum _{AB\in \alpha }\frac{c_{AB} }{\left(1+c_{A} \right)\left(1+c_{B} \right)}  \right)\prod _{D\in \alpha }\frac{\sigma _{D} }{\rho _{o} },    \end{eqnarray} 
defining $\Psi_\alpha $ as 

\begin{equation} \label{15} \Psi _{\alpha } =1+\sum _{AB\in \alpha }\gamma _{AB};\quad \gamma _{AB} =\frac{c_{AB} }{\left(1+c_{A} \right)\left(1+c_{B} \right)}.  \end{equation} 

Equation (\ref{14}) can be written in the concise form

\begin{equation} \label{16} \frac{\sigma _{\alpha } }{\rho _{o} } =\Psi _{\alpha } \prod _{A\in \alpha }\frac{\sigma _{A} }{\rho _{o} }  =\Psi _{\alpha } \prod _{A\in \alpha }\left(1+c_{A} \right).  \end{equation} 
In TPT1 all $c_{AB}$ = 0 resulting in $\Psi_\alpha  = 1$. 

Consider 3 cases: $\alpha $ = $\Gamma$, $\alpha$ = $\Gamma$ -- $A$ and $\alpha$ = $\Gamma - AB$
%
%
\begin{eqnarray} \label{17} &&\frac{\sigma _{\Gamma } }{\rho _{o} } =\frac{\rho }{\rho _{o} } =\Psi _{\Gamma } \prod _{A\in \Gamma }\left(1+c_{A} \right), \nonumber \\ 
&&\frac{\sigma _{\Gamma -A} }{\rho _{o} } =\Psi _{\Gamma -A} \prod _{A\in \Gamma -A}\left(1+c_{A} \right) , \nonumber\\ 
&&\frac{\sigma _{\Gamma -AB} }{\rho _{o} } =\Psi _{\Gamma -AB} \prod _{A\in \Gamma AB}\left(1+c_{A} \right),  \end{eqnarray} 
defining the monomer fraction, $X_o$,  fraction of spheres not bonded at site $A$, $X_A$, and the fraction of spheres not bonded at either site $A$ or $B$, $X_{AB}$, as

\begin{equation} \label{18} X_{o} =\frac{\rho _{o} }{\rho };\quad X_{A} =\frac{\sigma _{\Gamma -A} }{\rho } ;\quad X_{AB} =\frac{\sigma _{\Gamma -BC} }{\rho }.  \end{equation} 

From equations (\ref{17}) -- (\ref{18}) we obtain the fractions

\begin{eqnarray} \label{19} \frac{1}{X_{o} } &=&\Psi _{\Gamma } \prod _{A\in \Gamma }\left(1+c_{A} \right) , \nonumber \\ 
\frac{X_{A} }{X_{o} } &=&\Psi _{\Gamma -A} \prod _{A\in \Gamma -A}\left(1+c_{A} \right), \nonumber \\ 
\frac{X_{AB} }{X_{o} } &=&\Psi _{\Gamma -AB} \prod _{A\in \Gamma -AB}\left(1+c_{A} \right).  \end{eqnarray} 
Equations (\ref{19}) provide a general TPT2 solution for a pure fluid of associating spheres with a set of sites~$\Gamma$, subject to the approximation in equation (\ref{8}).

The total Helmholtz free energy is given by

\begin{equation} \label{20)} \frac{A-A_{hs} }{Nk_{\text{B}} T} =\ln \left(\frac{\rho _{o} }{\rho } \right)+\frac{Q}{\rho } +1-\frac{\Delta c_{}^{(o)} }{N},  \end{equation} 
where $A_{hs}$ is the free energy of the hard sphere reference fluid, $V$ is the system volume and $T$ is the absolute temperature. In TPT2, the $Q$ function is given by

\begin{equation} \label{21} \frac{Q}{\rho } =-1+\frac{1}{\rho } \sum_{\gamma \subset \Gamma \atop {\gamma \ne \emptyset }}c_{\gamma }  \sigma _{\Gamma -\gamma } =-1+\sum _{A\in \Gamma }\left(1-\chi _{A} \right) +\sum _{AB\in \Gamma }c_{AB} \frac{\Psi _{\Gamma -AB} }{\Psi _{\Gamma } } \chi _{A} \chi _{B}\,,   \end{equation} 

\begin{equation} \label{22} \chi _{A} =\frac{1}{1+c_{A}}.  \end{equation} 
Equation (\ref{21}) is as far as we can go without defining a specific form for $\Delta c^{(o)}$. 
Equations \ref{19} completely and concisely specify the associated state of the fluid, whereas the complete TPT2 solution \cite{10,11} requires the evaluation of recursion relations. The simple form of equations~(\ref{19}) will allow for wide applicability. We will refer to this approach as simplified second order perturbation theory, TPT2S. In section~3, TPT2S is applied to predict the effect of hydrogen bond cooperativity on the hydrogen bond structure in liquid water. 

\section{Application to hydrogen bond cooperativity in water}

In this section we demonstrate the accuracy of TPT2S for the description of hydrogen bond cooperativity in water. Marshall~\cite{12} developed a full TPT2 theory for water to describe positive hydrogen bond cooperativity. In this section, we compare TPT2S to this standard TPT2 solution.

Water is taken as a sphere with 4 association sites, 2 oxygen acceptors and 2 hydrogen donors, in the set $\Gamma =$ (O$_{1}$, O$_{2}$, O$_{1}$, H$_{2}$). The pairwise hydrogen bonding energy between an oxygen acceptor and hydrogen donor is $\varepsilon_{\text{OH}}$. However, it is well known~\cite{13} that water exhibits substantial hydrogen bond cooperativity (HBC). HBC is a cooperative effect resulting in the energetics of a given hydrogen bond interaction being dependent on the hydrogen bonding state of the molecule. Marshall~\cite{12} developed a simple treatment of this non-additivity by proposing that trimer chains in which the center molecule is bonded at a hydrogen and oxygen, HO-HO-HO, will exhibit a strengthening of the pairwise contribution. In a later paper, Marshall~\cite{14} also accounted for negative cooperativity for trimer chains in which the center molecule was bonded at either both hydrogens, HO-HH-OH, or both oxygen sites, OH-OO-HO. Accounting for negative cooperativity requires a resummation which we will not perform here. Hence, attention is restricted to the case of positive cooperativity in HO-HO-HO trimer clusters. Note, in Wertheim's multi-density formalism all trees of associated clusters are then created from the first order H-O and second order HO-HO-HO contributions. 

The energy of an HO-HO-HO trimer is given by

\begin{equation} \label{23} \varepsilon _{\text{HOHO}} =2\varepsilon _{\text{OH}} +\Delta \varepsilon _{\text{HOHO}} =\varepsilon _{\text{OH}} +\varepsilon _{\text{OH}+}.  \end{equation} 
The quantity $\Delta \varepsilon_{\text{HOHO}}$ is the cooperative contribution and $\varepsilon_\text{OH+}$ is the effective energy of the second hydrogen bond. The first and second order contributions to the fundamental graph sum are~\cite{12}

\begin{equation} \label{24} \frac{\Delta c_{I}^{\left(o\right)} }{V} =\frac{1}{2} \sum _{A\in \Gamma }\sum _{B\in \Gamma }\sigma _{\Gamma -A}   \sigma _{\Gamma -B} \Delta _{AB}\,,  \end{equation} 

\begin{equation} \label{25} \frac{\Delta c_{II}^{\left(o\right)} }{V} =\frac{1}{2} \sum _{A\in \Gamma }\sum _{B\in \Gamma }\sum _{C\in \Gamma }\sum _{D\in \Gamma }\sigma _{\Gamma -A}  \sigma _{\Gamma -B} \sigma _{\Gamma -CD} \Delta _{AC} \Delta _{BD}    \left(\delta _{CD} -1\right), \end{equation} 
where~\cite{12}

\begin{equation} \label{26} \Delta _{AB} =\kappa _{AB} g_{r} d^{3} f_{\text{OH}}\,,  \end{equation}

\begin{equation} \label{27} \delta _{CD} =\left\{\begin{array}{c} {\begin{array}{l} {\frac{f_{\text{OH}+} }{f_{\text{OH}} } \quad \text{if} \quad CD\in \text{OH}}\,, \\ {} \end{array}} \\ {1\quad \quad \text{Otherwise}\,,\quad \quad } \end{array}\right.  \end{equation}

\begin{equation} \label{28} f_{\text{OH}} =\exp \left(\frac{\varepsilon _{\text{OH}} }{k_{b} T} \right)-1 ;\quad f_{\text{OH}+} =\exp \left(\frac{\varepsilon _{\text{OH}+} }{k_{b} T} \right)-1. \end{equation} 

\begin{table}[!b]
\caption{Model parameters~\cite{12}.}
\label{table1}
\begin{center}
\begin{tabular}{|c|c|c|c|}
\hline 
$d$(\AA) & $\varepsilon_{\text{OH}} / k_{b}(K)$ & $\varepsilon_{\text{OH}+}/k_{b} (K)$ & $\kappa_{\text{OH}}$ \\ \hline 
3 & 1587.7 & 1873.5 & 0.015 \\ \hline 
\end{tabular}
\end{center}
\end{table}

\begin{figure}[!t]
	\begin{center}
		\includegraphics [width=0.78\textwidth]  {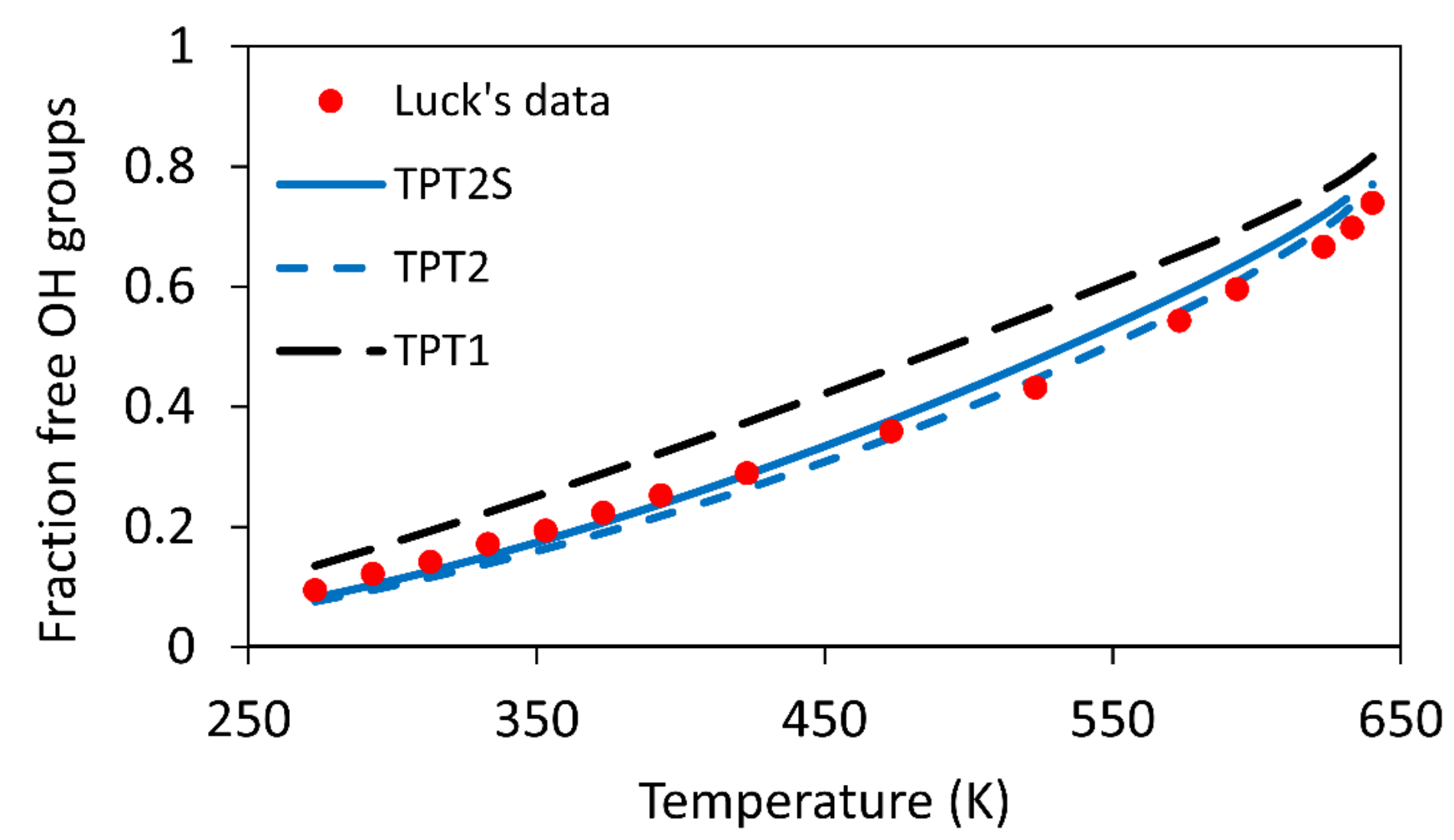}
		\caption{(Colour online) Comparison of model predictions to spectroscopic data~\cite{16} for the fraction of free OH groups in liquid water. TPT2 calculations were reproduced from Marshall~\cite{12}.} 
		\label{fig:Fig1}
	\end{center}
\end{figure}

\begin{figure}[!t]
	\begin{center}
		\includegraphics [width=0.78\textwidth]  {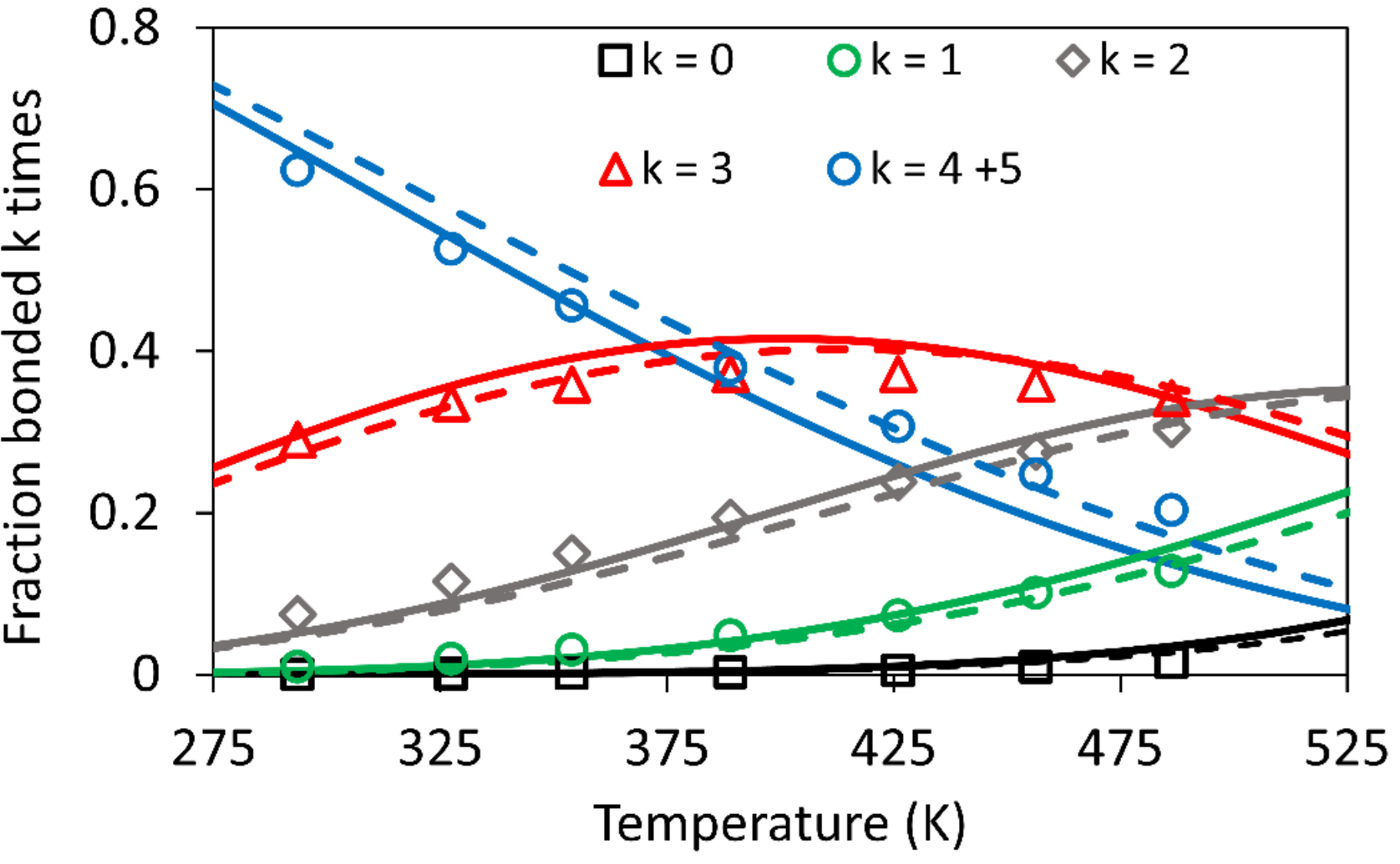}
		\caption{(Colour online) Comparison of TPT2S (solid curves), TPT2~\cite{12} (dashed curves) and molecular simulation data~\cite{17} using the $i$AOMEBA force field.} 
		\label{fig:Fig2}
	\end{center}
\end{figure}

Note, in the double sum over sites $C$ and $D$ in equation \ref{25}, it is understood that $\sigma _{\Gamma -CD} =\sigma _{\Gamma -DC}$. In equation (\ref{26}), $g_r$ is the contact value of the pair correlation function and $d$ is the sphere diameter. Equation (\ref{25}) relies on the linear superposition of the triplet correlation function. In the application to hydrogen bond cooperativity, TPT2 is applied to correct energetic contributions to the free energy, not structural effects. 

For conical square well association sites, the bond volume $\kappa_{AB}$ is given by \cite{15}

\begin{equation} \label{29} \kappa _{AB} =\piup \left(1-\cos \theta \right)^{2} \left(\frac{r_{c} }{d} -1\right). \end{equation} 

Conical square well association sites assume that if the centers of two spheres are separated by a distance $r<r_c$ and the angles $\theta_A<\theta_c$ and $\theta_B<\theta_c$, the two association sites are considered bonded. This specific form is the result of an approximation of the integral of the reference correlation function over the bond volume.

 From equations (\ref{6}), (\ref{24})--(\ref{25})

\begin{eqnarray} \label{30} c_{A} &=&\sum _{B\in \Gamma }\sigma _{\Gamma -B} \Delta _{AB}  +\sum _{B\in \Gamma }\sum _{C\in \Gamma }\sum _{D\in \Gamma }\sigma _{\Gamma -CD}  \sigma _{\Gamma -B} \Delta _{AC} \Delta _{BD}   \left(\delta _{CD} -1\right),\nonumber\\ 
c_{CD} &=&\sum _{A\in \Gamma }\sum _{B\in \Gamma }\sigma _{\Gamma -A} \sigma _{\Gamma -B} \Delta _{AC} \Delta _{BD} \left(\delta _{CD} -1\right).   \end{eqnarray} 
Enforcing the following relations among unbonded fractions

\begin{equation} \label{31} \begin{array}{l} {X_{\text{O}_{1} } =X_{\text{O}_{2} } =X_{\text{H}_{1} } =X_{\text{H}_{2} } =X_{\text{H}} }\,, \\ {} \\ {X_{\text{O}_{1} \text{H}_{1} } =X_{\text{O}_{1} \text{H}_{2} } =X_{\text{O}_{2} \text{H}_{1} } =X_{\text{O}_{2} \text{H}_{2} } =X_{\text{OH}}\,, } \end{array} \end{equation} 
 equations (\ref{30}) can be further simplified,

\begin{equation} \label{32} \begin{array}{l} {c_{\text{H}} =2\rho X_{\text{H}} \Delta _{\text{OH}} +8\rho ^{2} X_{\text{H}} X_{\text{OH}} \Delta _{\text{OH}}^{2} \left(\delta _{\text{OH}} -1\right)}, \\ {} \\ {c_{\text{OH}} =4\rho ^{2} X_{\text{H}}^{2} \Delta _{\text{OH}}^{2} \left(\delta _{\text{OH}} -1\right)}, \\ {} \\ {c_{\text{O}_{1} \text{O}_{2} } =c_{\text{H}_{1} \text{H}_{2} } =0}. \end{array} \end{equation} 

 All relations in this section have been entirely consistent with the original TPT2 approach. Now, we employ TPT2S to obtain the solution for the hydrogen bonding state of the system.

\begin{eqnarray} \label{33)} \Psi _{\Gamma } &=&1+4\frac{c_{AB} }{\left(1+c_{A} \right)\left(1+c_{B} \right)}, \nonumber\\ 
\Psi _{\Gamma -\text{H}} &=&1+2\frac{c_{AB} }{\left(1+c_{A} \right)\left(1+c_{B} \right)}, \nonumber \\ 
\Psi _{\Gamma -\text{OH}} &=&1+\frac{c_{AB} }{\left(1+c_{A} \right)\left(1+c_{B} \right)}. 
 \end{eqnarray} 
The required bonding fractions are then obtained through equations (\ref{19}).  So, once the $c_\text{H}$ and $c_{\text{OH}}$ have been calculated from the fundamental graph sum, application of TPT2S is trivial. 

 The model parameters were estimated through a combination of direct measurement and quantum mechanical calculation. We refer the reader to the original publication~\cite{12} for additional details. The parameters are summarized in table~\ref{table1}.

Figure~\ref{fig:Fig1} compares model predictions for the fraction of free OH groups (${X_\text{H}}$) to the spectroscopic data of Luck \cite{16}. We include predictions from TPT1, TPT2S, as well as the full TPT2 solution of Marshall \cite{12}. As can be seen, TPT2S and TPT2 give similiar predictions while TPT1 is in significant error. TPT2S predicts a slightly larger $XH$ as compared to TPT2, meaning TPT2S predicts a slightly lower degree of hydrogen bonding. This is consistent with expectation due to the neglect of contributions which contain the product $c_{\text{OH}} c_{\text{OH}}$ in TPT2S.

 Figure~\ref{fig:Fig2} compares model predictions using both TPT2 and TPT2S to the molecular simulations of Fouad {et al}.~\cite{17} for the fraction of molecules hydrogen bonded $k$ times in liquid water. Both TPT2 and TPT2S give a good representation of the simulation data. Further, the predictions of both approaches for the fractions $k = 0 - 3$ are nearly identical. There is a small difference in the fraction of molecules bonded 4 times (\textit{k} = 4), where TPT2S predicts a lower fraction of fully hydrogen bonded water as compared to TPT2. However, the two theories give practically the same overall result. This demonstrates the utility of the assumption in equation~(\ref{8}).

\section{Conclusions and future work}

 A simplified TPT2 solution has been obtained for a pure fluid of associating spheres with any number of association sites. The solution is general in that it allows for inclusion of a variety of second order effects including steric hindrance between association sites, ring formation, double bonding and hydrogen bond cooperativity. The hydrogen bonding state of any species in TPT2 can be obtained through solution of equations (\ref{19}). The simplicity of this result will allow for a general implementation of the theory in which each pair of association sites can participate in these second order effects. This concise form was made possible by the assumption in equation (\ref{8}). It was demonstrated that the simplified second order theory (TPT2S) gave nearly identical predictions to the rigorous TPT2 theory for the prediction of liquid water hydrogen bonding structure.

\ukrainianpart

\title{Проста термодинамічна теорія збурень другого порядку для асоціативних плинів
}
\author[Б. Д. Маршал]{Б. Д. Маршал}
\address{Дослідницька група ExxonMobil, 1545 Рт. 22 Іст, Аннандейл, NJ 08801}

\makeukrtitle
	
	\begin{abstract}
		Пропонується апроксимація в межах термодинамічної теорії збурень Вертгайма другого порядку, яка дозволяє отримати загальний розв'язок для компонент плинів з довільною кількістю та функціональністю асоціативних центрів. Розв'язок є замкнутим, стислим та загальним для всіх явищ другого порядку, таких як формування кілець, стеричних перешкод та функціональностей асоціативних центрів. Дієвість даного методу підтверджується шляхом порівняння з даними щодо структури водневих зв'язків у воді.
		
\keywords термодинаміка, класична статистична механіка, самозбірка
		
	\end{abstract}
\lastpage
\end{document}